# Realization of practical level current densities in $Sr_{0.6}K_{0.4}Fe_2As_2$ tape conductors for high-field applications

Xianping Zhang,[1, a)] Chao Yao,[1, a)] He Lin,[1] Yao Cai,[2] Zhen Chen,[2] Jianqi Li,[2] Chiheng Dong,[1] Qianjun Zhang,[1] Dongliang Wang,[1] Yanwei Ma,[1, b)] Hidetoshi Oguro,[3] Satoshi Awaji,[3] Kazuo Watanabe[3]

[1] *Key Laboratory of Applied Superconductivity, Institute of Electrical Engineering, Chinese Academy of Sciences, Beijing 100190, China*

[2] *Beijing National Laboratory for Condensed Matter Physics, Institute of Physics, Chinese Academy of Sciences, Beijing 100190, China*

[3] *High Field Laboratory for Superconducting Materials, Institute for Materials Research, Tohoku University, Sendai 980-8577, Japan*

122 type pnictide superconductors are of particular interest for high-field applications because of their large upper critical fields $H_{c2}$ (> 100 T) and low anisotropy γ (<2). Successful magnet applications require fabrication of polycrystalline superconducting wires that exhibit large critical current density $J_c$, which is limited by poor grain coupling and weak-link behavior at grain boundaries. Here we report our recent achievement in the developing $Sr_{0.6}K_{0.4}Fe_2As_2$ tapes with transport $J_c$ up to 0.1 MA/cm$^2$ at 10 T and 4.2 K. This value is by far the highest ever recorded for iron based superconducting wires and has surpassed the threshold for practical application. The synergy effects of enhanced grain connectivity, alleviation of the weak-link behavior at grain boundaries, and the strong intrinsic pinning characteristics led to the superior $J_c$ performance exhibited in our samples. This advanced $J_c$ result opens up the possibility for iron-pnictide superconducting wires to win the race in high-field magnet applications.

---

a) X. Zhang and C. Yao contributed equally to this work.

b) Author to whom any correspondence should be addressed. Electronic mail: ywma@mail.iee.ac.cn

Discovered as a member of iron-based superconductor family in 2008,[1] the 122 type iron-pnictides ($AeFe_2As_2$, Ae = alkali or alkali earth elements) are particularly interest for high-field magnet applications, because of their superior transport $J_c$ over 1 MA/cm$^2$ achieved in thin films, small $H_{c2}$ anisotropy < 2 and high upper critical fields $H_{c2}$ above 100 T observed in single crystals.[2,3] For large-scale applications, superconducting wires with the ability to conduct high supercurrents are particularly desirable. To date, iron-pnictide wires are mainly fabricated based on the powder-in-tube (PIT) method.[4] However, it is especially challenging to make high-performance iron-pnictide wires, due to the fact that the transport current is limited by poor grain coupling and high-angle grain boundaries (GBs), which are ineluctably introduced into the superconducting core during the mechanical deformation process of the PIT method.

The defects such as cracks, pores and impurity phases in the superconducting core of wires can block the current paths and thus degrade the value of transport current density.[4] In order to improve the connectivity between grains, *ex situ* PIT technique was employed to reduce pores formed during heat treatments,[5,6] metal additives were used to strengthen grain coupling,[7,8] and mechanical densification process was optimized to increase mass density[9-14] in the fabrication of $Ba_{1-x}K_xFe_2As_2$ (Ba-122) and $Sr_{1-x}K_xFe_2As_2$ (Sr-122) wires and tapes. On the other hand, the weak-link behavior at misoriented GBs can deteriorate supercurrents regardless of the mass density of materials[15]. Thus, we can reasonably expect high transport $J_c$ in 122 iron-pnictide tapes with *c*-axis aligned grains similar to those reported for Bi-2223 tapes.[16] Therefore, some endeavors have been made to achieve *c*-axis texture in 122 iron-pnictide tapes by flat rolling or cold pressing process, and their results show improvements on transport $J_c$ due to the strengthened grain orientations.[17-19]

In the past two years, significant progresses towards high-performance iron-pnictide conductors were made, and the $J_c$ values of Sr-122 and Ba-122 wires and tapes have rapidly increased above $10^4$ A/cm$^2$ at 4.2 K and 10 T by various deformation techniques.[11-14, 18] Nevertheless, the threshold value for practical application, which is generally accepted as $10^5$ A/cm$^2$ at 10 T, has not been achieved.

Here we report our accomplishment in developing $Sr_{0.6}K_{0.4}Fe_2As_2$ tapes with transport $J_c$ up to $10^5$ A/cm$^2$ at 10 T and 4.2 K by a hot pressing (HP) technique. This achievement sets a record in $J_c$ value for iron-pnictide wires and this is considered as a technical breakthrough in reaching the $J_c$ level desired for practical applications.

Ag-clad $Sr_{0.6}K_{0.4}Fe_2As_2$ tapes with Sn addition were fabricated by the *ex-situ* PIT method. Sr fillings, K pieces, and Fe and As powders were used as staring materials with the nominal composition of $Sr_{0.6}K_{0.5}Fe_2As_{2.05}$. The starting materials were mixed and ground by ball-milling method under argon atmosphere. Then the milled powders were heat treated at 900 ℃ for 35 h. The sintered bulk was ground into powders and mixed with 5 wt% Sn. The precursors were packed into Ag tubes with OD 8 mm and ID 5 mm. These tubes were sealed and then cold worked into tapes (about 0.4 mm in thickness) by drawing and flat rolling. We also fabricated 7- and 19-filamentary tapes using a similar process. On the other hand, for the hot pressed samples, the flat rolled mono- and multifilamentary tapes were sandwiched between two pieces of metal sheets and pressed at ~30 MPa at 850 ℃ for 30 min. The transport current $I_c$ at 4.2 K and its magnetic field dependence were measured by the standard four-probe method with a criterion of 1 $\mu$V/cm.

Fig.1(a) shows the magnetic field dependence of the $J_c$ of the pressed Sr-122 tapes along with the $J_c$ values of flat rolled tapes. For the flat rolled tapes, $J_c$ achieved a maximum value of $3\times10^4$ A/cm$^2$ at 10 T, 4.2 K. After the pressing, the $J_c$ of Sr-122 sample is shown to be over $10^5$ A/cm$^2$ in 10 T at 4.2 K, and still remain $8.4\times10^4$ A/cm$^2$ up to 14 T. Notably, the $J_c$ values obtained here is by far the highest ever reported for the iron-based wires and tapes. In order to further investigate their potential for practical applications, in which wires and tapes with multifilamentary structure are more favorable due to the stable superconducting properties against magnetic flux jumps and thermal quenching, we fabricated multifilamentary Sr-122 tapes by flat rolling and pressing, respectively. As shown in Fig.1 (a) and (b), after pressing, even the 7- and 19-filamentary tapes still sustain $J_c$ as high as $6.1\times10^4$ A/cm$^2$ and $3.5\times10^4$ A/cm$^2$ at 10 T and 4.2 K, respectively, which are the best record in multifilamentary pnictide wires at present. Fig.1(b) presents the highest $J_c$ data recorded so far for both

Sr-122 and Ba-122 mono- and multifilamentary tapes. To compare, the $J_c$ data for conventional NbTi and $Nb_3Sn$ superconducting wires and the PIT-processed $MgB_2$ wires were also included. An interesting point is the comparison of the superconducting properties of the present tapes with $MgB_2$, NbTi and $Nb_3Sn$. Clearly, the pressed 122 iron-pnictide tapes have superior $J_c$ than $MgB_2$ and NbTi in field region over 10 T. As seen from the figure, it is a common feature for 122 iron-pnictide tapes that the $J_c$ show very weak field dependence, which is a desirable character for high-field applications.

XRD patterns of the precursor and the tapes processed by rolling and pressing were shown in Fig.2(a). Obviously, both of the tape samples exhibit a well-defined $ThCr_2Si_2$-type crystalline structure except for the Ag peak which is induced by Ag sheath. The XRD data indicate that a single-phase $Sr_{1-x}K_xFe_2As_2$ superconductor was obtained. Compared with the randomly oriented precursor, stronger relative intensity of (*00l*) peaks was observed in both rolled and pressed tapes, indicating a good *c*-axis texture of Sr-122 grains. It is believed that the rolling and pressing process promotes grain alignment in *c*-axis, because the planar Sr-122 crystals are prone to rotate along the tape surface by the deformation force. The degree of the grain alignment can be estimated by the Lotgering method[20] with $F=(\rho-\rho_0)/(1-\rho_0)$, where $\rho=\Sigma I(00l)/\Sigma I(hkl)$, $\rho_0=\Sigma I_0(00l)/\Sigma I_0(hkl)$. $I$ and $I_0$ are the intensities of each reflection peak (*hkl*) for the textured and randomly oriented samples, respectively. The *c* axis orientation factors $F$ for rolled and pressed samples are 0.45 and 0.52, respectively, showing an improvement on grain texture after pressing.

From the *M-T* curve of the pressed samples in Fig.2(b), we observe a temperature-independent magnetic moment at low temperatures and a sharp superconducting transition, underlining the high quality of the samples. The pressed samples show a stronger diamagnetic signal than the rolled one, indicating that a better superconducting phase was formed after the pressing. As shown in Fig.2(c), the pressed samples have $T_c^{onset}$ and $T_c^{zero}$ about 35.4 K and 34.7 K at 0 T, respectively. The sharp resistivity transition demonstrates high degree of electromagnetic homogeneity of the pressed samples. Fig.2(d) shows the temperature dependence of upper critical field $H_{c2}$ and irreversibility field $H_{irr}$. Here we use 90% and 10% points of normal state resistivity on the resistive transition curves in Fig.2(c) to estimate the $H_{c2}$ and $H_{irr}$. $H_{c2}(0)$ is estimated to be about 199 T and 119 T for $H$ parallel to $ab$-plane and $c$-axis respectively. These results are comparable to that measured on the K-doped Ba-122 single crystal.[21] At the same time, small anisotropy of $H_{c2}$ ($\gamma=H_{c2}^{H//ab}/H_{c2}^{H//c}$ ~1.68) was observed, which is beneficial to magnets application. In addition, it should be realized that the $H_{irr}$ determines the regime in the $T$-$H$ phase diagram where this material can be operated.[2] Therefore, the large $H_{irr}$ exhibited in the pressed samples signifies the suitability of this material for high-field applications.

Figures 3(a) and 3(b) are the Arrhenius plots of resistivity as Ln $\rho$ vs. $T^{-1}$ for both directions measured in static fields up to 13 T, showing the thermally activated behavior of resistance. Fig. 3(c) presents the pinning potential $U_0/K_B$[22] obtained from the slop of the linear portion of the curves in Figures 3(a) and 3(b). It can be seen that $U_0(H)$ follows a power law [i.e., $U_0(H) \sim H^{-\alpha}$] for both directions. On the other hand, the activation energy exhibits constant power-law dependences on magnetic field, with $\alpha = 0.26$ in the case when field is parallel to the $c$-axis. This scale value is comparable to that of Ba-122 single crystal, which is almost field independent.[23] The best fit of the experimental data yields value of the pinning potential $U_0/K_B \sim 7175$ K in 0.5 T for the $H$ parallel to $c$-axis which is comparable to that reported for $MgB_2$[24] (8700 K) and significantly larger than that reported for Bi-2212[22] (700 K) and Bi-2223[25] (650 K) measured under the same field, 0.5 T. Furthermore, the pressed Sr-122 tapes exhibit larger pinning potential than $MgB_2$ and YBCO[26] in the field of $H$>2 T, and larger than BSCCO in the whole magnetic field. As illustrated in Fig.3 (d), the volume pinning force $F_p$ of the pressed Sr-122 tapes exceeded the value of $MgB_2$[27] and NbTi[28] above 10T, and may have a crossover with that of $Nb_3Sn$[29] at around 15 T. This confirms that the pressed Sr-122 tapes have strong pinning characteristics in high magnetic fields, and is in agreement with the very weak $J_c$-field dependence in Fig.1 and the high values of the $H_{irr}$ in Fig.2(d).

The typical scanning electron microscopy (SEM) images of rolled and pressed Sr-122 samples are exhibited in Fig.4. The observation of images was performed on the core surface after peeling off the Ag sheath. Compared to the loose structure of the rolled sample in Fig.4(a), a much denser Sr-122 core is shown in Fig.4(b) for the pressed tapes. On the other hand, the presence of uniformly textured grains along the tape axis is clearly observed from Fig.4(c). More importantly, as shown in the enlarged part in Fig.4(c), grain bending can be observed throughout the pressed sample, which is thought to be caused by the softness of grains during hot deformation. The bending of grains can prevent the formation of cracks and improve the grain coupling effectively, inducing substantial enhancement in grain linkages. This is similar to the case of Ag/Bi-2223 tapes, in which hot pressing is also very effective in healing pores and cracks.[30, 31] As a result, in our hot pressed Sr-122 samples, we did not observe residual cracks as in the cold pressed Ba-122 tapes,[12] suggesting that hot deformation can help in eliminating cracks effectively while densifying the superconducting cores.

In order to further investigate the microstructure of our pressed Sr-122 tapes, the superconducting core processed by focused ion beam (FIB) technique was submitted to transmission electron microscopy (TEM) examination. As shown in Fig.5, the good homogeneity of our tapes has been further confirmed by energy dispersive X-ray spectroscopy (EDS) mapping based on TEM, which demonstrates that the element distribution of Sr-122 phase is homogeneously dispersed in the superconducting core. With high resolution TEM (HRTEM) imaging, we studied the GBs of Sr-122 grains. Fig.6(a) shows a typical image of three Sr-122 grains (marked as A, B and C) in longitudinal view of the hot pressed tapes, the GBs between the grains are very clean, and no amorphous contrast is observed, indicating the good Sr-122 phase purity and the well-connected Sr-122 grains to each other. Fig.6(b) presents two Sr-122 grains (marked as D and E) with a small misorientation angle $< 3^o$, which can clearly be seen in the fast fourier transformation of the HRTEM image in the inset. Enlarged image of the GB between grain A and B is shown in Fig.6(c), selected area electron diffraction (SAED) shown in the inset confirmed that both the lattice fringe in A and B grains refer to the [100] direction, and the [100]-tilt misorientaion angle is $8.4^o$. Fig.6(d) shows the GB between other two grains (marked as F and G) with a small misorientaion angle of 7.6$^o$. In the pressed samples, lots of such low-angle GBs can be observed between the Sr-122 grains. Therefore, TEM investigations reveal the good homogeneity and clean GBs with low misorientaion angle inside our tapes. It is reported that the transport $J_c$ will not suffer much depression when it run across GBs with small misoriented angles < 9$^o$ in Ba-122 bicrystal[32], so such low-angle GBs are advantagous to high transport $J_c$.

Our experimental results, including superconducting properties and microstructure analysis, reveal that the superior $J_c$ achieved in our Sr-122 tapes may be attributed to the combination of significantly improved grain connectivity, good grain texture, and strong pinning characteristics. Firstly, in rolled Sr-122 tapes, the large deformation force may produce residual microcracks that cannot be healed by heat treatment. In contrast, as shown in Figures 4(b) and (c), the HP technique can make the grains more flexible to couple with each other without producing a large number of crashed grains. Therefore, our fabrication process results in an excellent balance between the mechanical deformation induced densification and the cracks caused by deformation pressure, which is the key point to achieve such superior $J_c$. In addition, the pressure applied during sintering process can improve the formation of the superconducting phase, thus increasing the phase quality, as evident in the M-T plot shown in Fig.2(b). The segregation of secondary phases around GBs of the pnictide superconductor is frequently cited as the limiting factor for $J_c$ because of the concomitant degradation in intergranular connectivity[33, 34]. For our hot pressed tapes, the HRTEM results images displayed in Figures 6(a-d) clearly show that GBs is very clean. Obviously, compact superconducting core, elimination of the cracks, and clean grain boundaries attribute to the good grain connectivity in the pressed Sr-122 tapes. Secondly, the grain alignment confirmed by the XRD and SEM analysis in Figures 2(a) and 4(c) is beneficial to the high transport $J_c$ obtained in our hot pressed tapes, since lots of low-angle GBs were observed in the superconducting core by HRTEM investigation. Thirdly, for high-field magnet applications, large current carrying capability，strong pinning characteristics and low superconductivity anisotropy are desirable. As supported by the larger pinning potential $U_0$ of >7000 K at 0.5 T with small α of ~0.26 for *H//c*-axis, larger transport critical currents in high fields were achieved in our pressed tapes. Therefore, by combining the deformation improved microstructure with the high intrinsic flux pining characteristic, the Sr-122 tapes prepared in this work show an excellent transport $J_c$ of 0.1 MA/cm$^2$ at 10 T, which is well on the level of practical applications.

**Acknowledgments**

The authors thank Beihai Ma, Zhixiang Shi, Mingliang Tian, Haihu Wen, Huan Yang and Haitao Zhang for help and useful suggestion. This work is partially supported by the National '973' Program (grant No. 2011CBA00105), the National Natural Science Foundation of China (grant Nos. 51025726, 51172230, 51202243 and 51320105015), and the Beijing Municipal Natural Science Foundation (No. 2122056).

# Captions

FIG. 1  (a) The field dependence of transport $J_c$ values at 4.2 K of flat rolled and pressed Sr-122 tapes obtained in this experiment. (b) The $J_c$ properties of the 122 tapes compared to commercial NbTi, Nb$_3$Sn and MgB$_2$ wires. The magnetic field was applied parallel to the tape surface.

FIG. 2 (a) The XRD patterns of the precursor, flat rolled and pressed samples. (b) Magnetization as a function of temperature for the flat rolled and pressed samples. (c) Temperature dependent resistivity for pressed samples measured in different magnetic fields. (d) Upper critical field $H_{c2}$(T) and irreversibility field $H_{irr}$(T) of the pressed samples. All data were obtained on the small pieces after peeling off the Ag sheath.

FIG. 3  (a) Arrhenius plots of $\rho$ at various magnetic fields parallel to the *ab*-plane and (b) *c*-axis. The activation energy $U_0$ is given by the slopes from linear fitting. (c) Magnetic field dependence of the activation energy $U_0$. (d) The flux pinning force ($F_p$) at 4.2 K plotted as a function of applied magnetic fields. Transport $J_c$ was used in the calculation of $F_p$ by the formula $F_p=J_c\times B$. The $F_p$ data at 4.2 K for MgB$_2$, NbTi, Nb$_3$Sn were also included for comparison.

FIG. 4  Microstructure of the core region of Sr-122 tapes investigated by SEM. (a) Flat rolled samples viewed from the tape surface direction. Some voids and cracks were indicated by arrows or circles. (b) Pressed samples viewed from the tape surface direction. (c) Pressed samples viewed from the longitudinal cross-sections. SEM analysis was performed after peeling off the Ag-sheath.

FIG. 5 TEM-EDS mapping images of the core region of pressed Sr-122 tapes.

FIG. 6 HRTEM study on the grain boundaries (GBs) of pressed Sr-122 tapes. (a)

HRTEM observation of the Sr-122 superconducting core in longitudinal direction of the tape, showing clean GBs of three grains (marked as A, B and C). (b) HRTEM image showing a very small misorientaion angle $< 3^\circ$ of two grains (marked as D and E). (c) Enlarged image of the GB between grain A and B with a low misorientaion angle of $8.4^\circ$. (d) HRTEM image of the GB between grain F and G with a low $c$-axis tilt misorientaion angle of $7.6^\circ$. The insets in (a), (b), and (d) are the fast fourier transformation of the HRTEM images. The inset in (c) is the selected-area electron diffraction image of the grains.

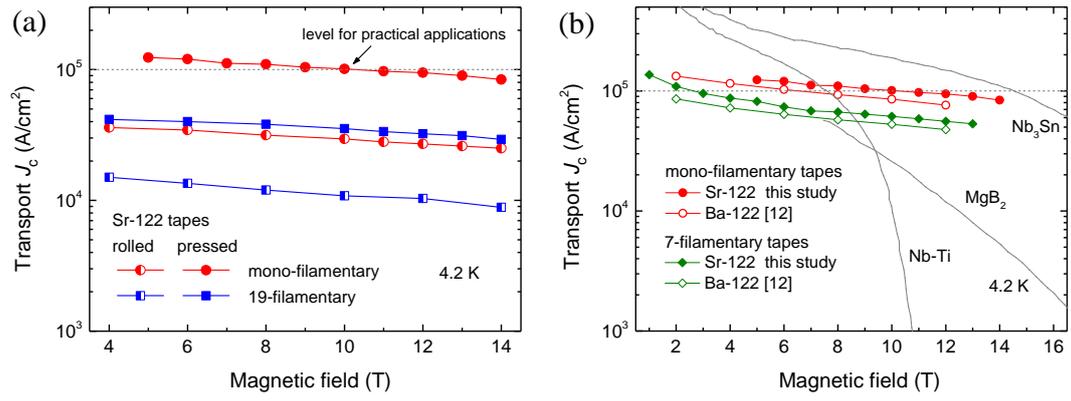

Figure. 1

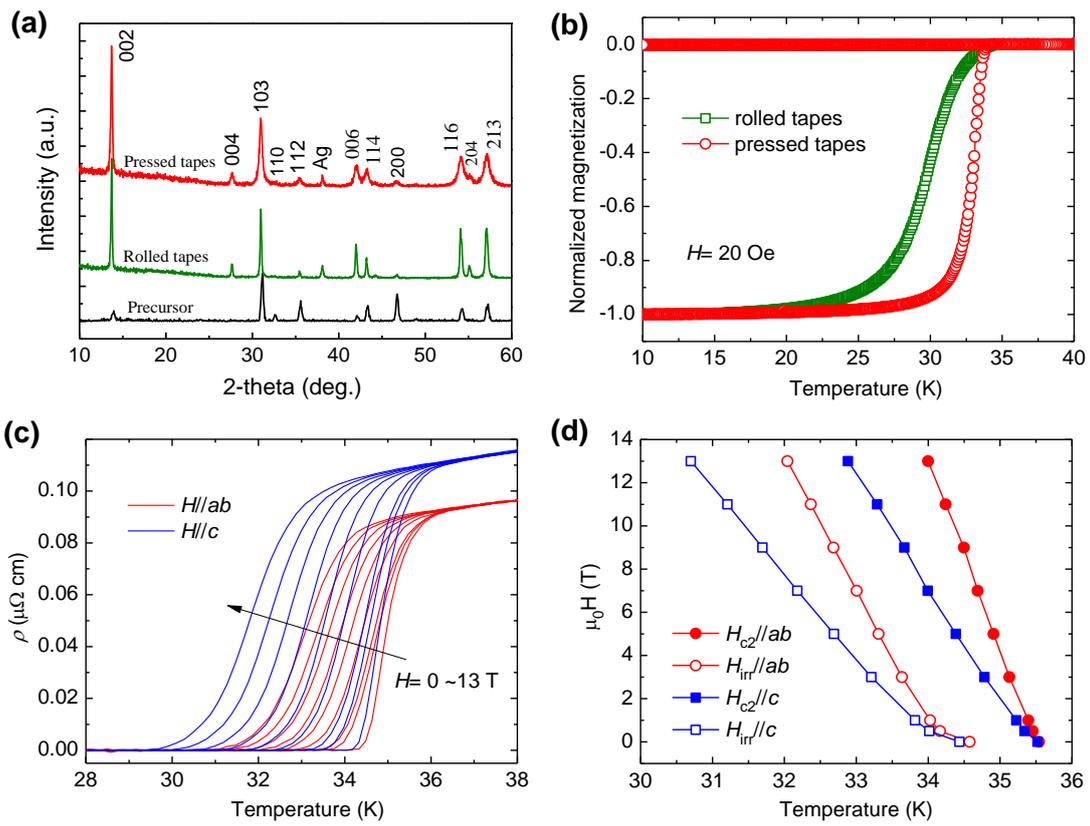

Figure. 2

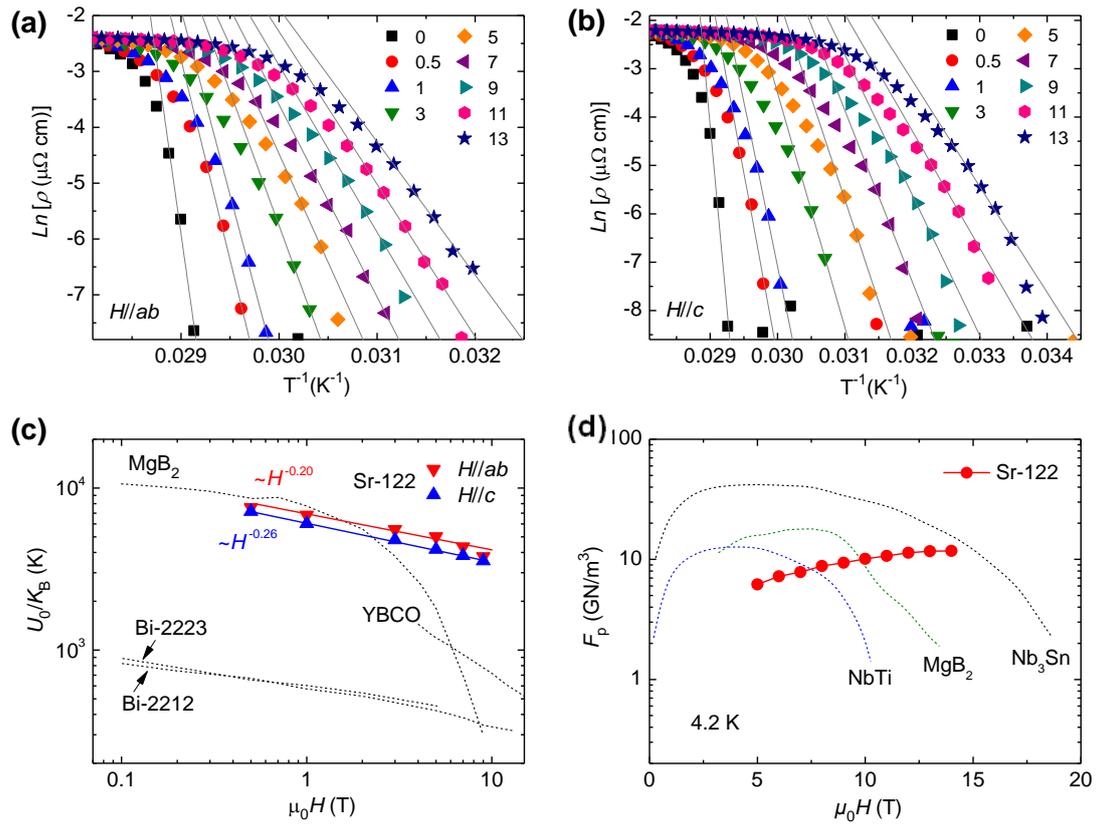

Figure. 3

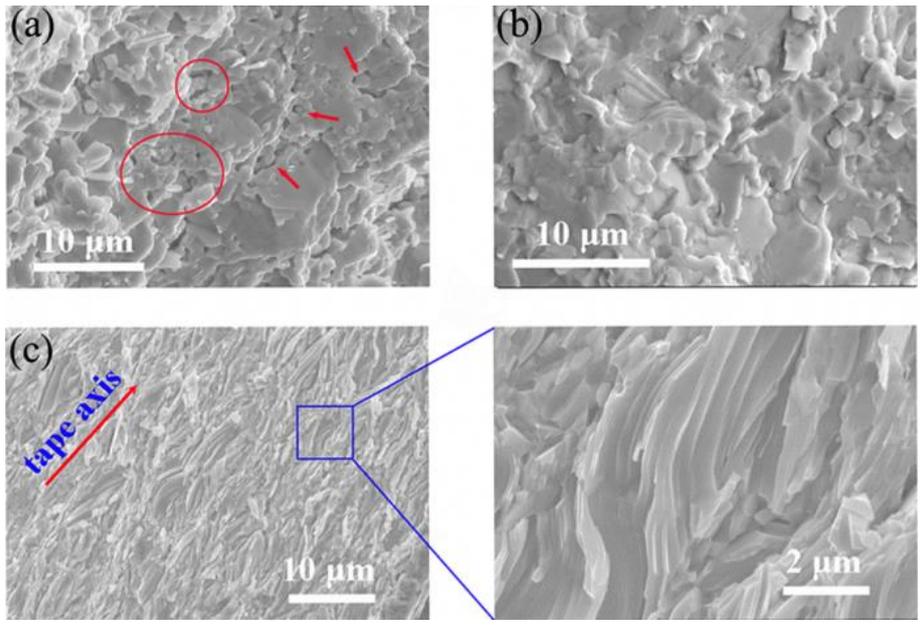

Figure. 4

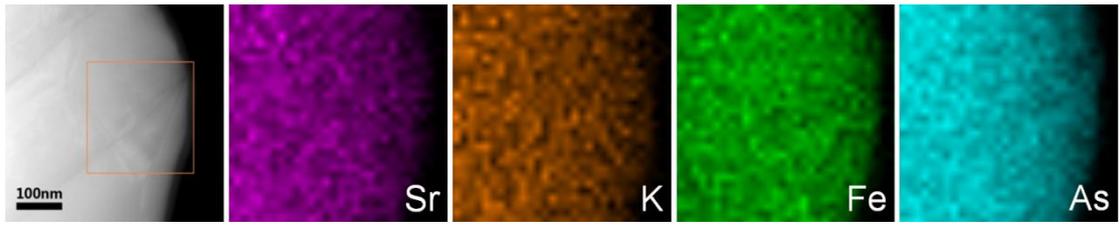

Figure. 5

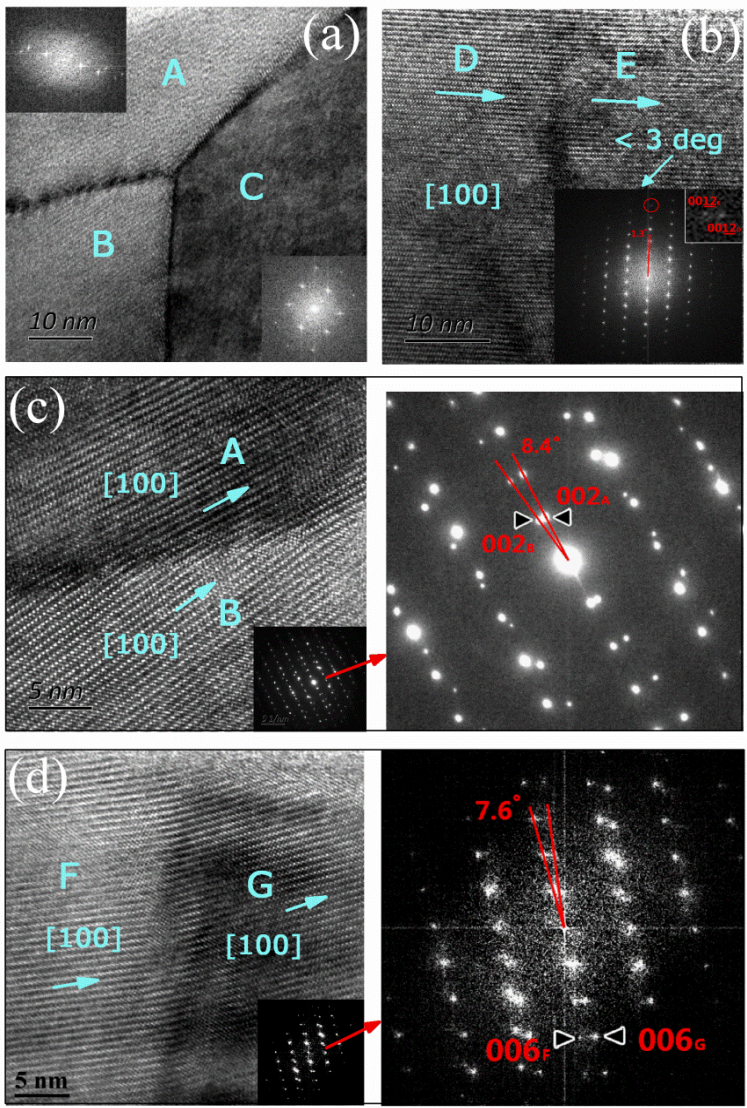

Figure. 6